\documentclass[prx,aps,twocolumn,noshowpacs,shortbibliography,noshowkeys,superscriptaddress]{revtex4-1}
\usepackage{amsfonts}
\usepackage{amsmath}
\usepackage{amssymb}
\usepackage{graphicx}

\usepackage{textcomp}%
\newcommand{\vv}[1]{\mathbf{#1}}

\usepackage{color}

\begin{document}
\def\neel{Institut N\'{e}el, Universit\'{e} Grenoble Alpes - CNRS:UPR2940, 38042 Grenoble, France}
\def\ilm{Institut Lumi\`{e}re Mati\`{e}re, UMR5306, CNRS - Universit\'{e} Claude Bernard Lyon 1, 69622 Villeurbanne, France}
\author{Laure Mercier de L\'{e}pinay}
\affiliation{\neel}
\author{Benjamin Pigeau}
\affiliation{\neel}
\author{Benjamin Besga}
\affiliation{\neel}
\author{Olivier Arcizet}
\affiliation{\neel}
\email{olivier.arcizet@neel.cnrs.fr}

\title{
Eigenmode orthogonality breaking and deviation from the fluctuation dissipation relation in rotationally dressed nanomechanical oscillators
}

\begin{abstract}
{\bf
The ultimate sensitivities achieved in force or mass sensing are limited by the employed nanomechanical probes thermal noise. Its proper understanding is critical for ultimate operation and any deviation from the underlying fluctuation dissipation theorem should be carefully inspected. Here we investigate an ultrasensitive vectorial force-field sensor, a singly clamped nanowire oscillating along two quasi frequency degenerated transverse directions. Immersing the nanowire in a non-conservative optical force field causes dramatic modifications of its thermal noise and driven dynamics. In regions of strong vorticity, eigenmodes orientations are distorted and lose their initial orthogonality. Thermal noise spectra strongly deviate from the normal mode expansion and presents an anomalous excess of noise violating the fluctuation dissipation theorem. Our model quantitatively accounts for all observations and underlines the role of non-axial response when patching the fluctuation dissipation relation. These results reveal the intriguing properties of thermal fluctuations in multimode nano-optomechanical systems and the subtleties appearing when performing thermal noise thermometry in such systems. They are also valid in any non-symmetrically coupled dual systems.
\\}
\end{abstract}
\maketitle

{\it Introduction--}
The sensitivity of any sensor based on mechanical oscillators is intrinsically limited by its thermal noise, a random position fluctuation which can mask the signal under investigation. Hence, understanding and reducing thermal noise is a permanent objective in force, position or metric sensing when approaching ultimate sensitivities.\\
The recent advent of nanomechanical oscillators \cite{Cleland2003,Ekinci2005,Schwab2005} has boosted the force sensitivity by orders of magnitudes \cite{Moser2013}, enabling single electron imaging \cite{Rugar2004} or ultimate mass sensing \cite{Jensen2008,Chaste2012,Sage2015},  opening perspectives in both fundamental and applied science \cite{Arcizet2011,Peddibhotla2013, Yeo2014, Montinaro2014,Pigeau2015}.
Their ultralow-mass causes a significant increase of thermal noise, that can spread over distances  approaching their intrinsic dimensions and makes the nanoresonator even more sensitive to environmental inhomogeneities which manifest themselves as external force field gradients. They can be mapped by tracking modifications of the probe nanomechanical properties as standardly used in atomic force microscopes \cite{Binnig1986} which probe force gradients along the oscillating direction of the force probe, either normal or parallel \cite{Karrai1995} to the sample. When the force probe can vibrate along several directions in space, vectorial force field imaging becomes accessible \cite{Gloppe2014,MercierdeLepinay2016,Rossi2016}, conveying a great physical richness. In particular the role of non-conservative force fields -in the sense that they do not derive from a potential energy or equivalently that the force field presents vorticity- can now be inspected. \\
This work investigates the thermal noise and dynamics of a 2D nanomechanical force probe, a singly clamped  suspended nanowire oscillating along both transverse directions \cite{Siria2012,Gloppe2014,Nichol2008,Gil-Santos2010,Ramos2012, MercierdeLepinay2015,Pigeau2015,Cadeddu2016,Rossi2016,MercierdeLepinay2016} and immersed in a tunable rotational force field. We report on the observation of the warping of the eigenmodes basis, breaking its original orthogonality, on an alteration of their thermal noise spectra and on a violation of the fluctuation dissipation theorem (FDT) in its original formulation. All these rather intriguing observations are quantitatively explained by our model which also conveys a generic methodology to correctly describe the fluctuations and dynamical properties of dual resonators strongly coupled by non-conservative mechanisms, which are often encountered in physics.\\
\begin{figure}[t!]
\begin{center}
\includegraphics[width=0.99\linewidth]{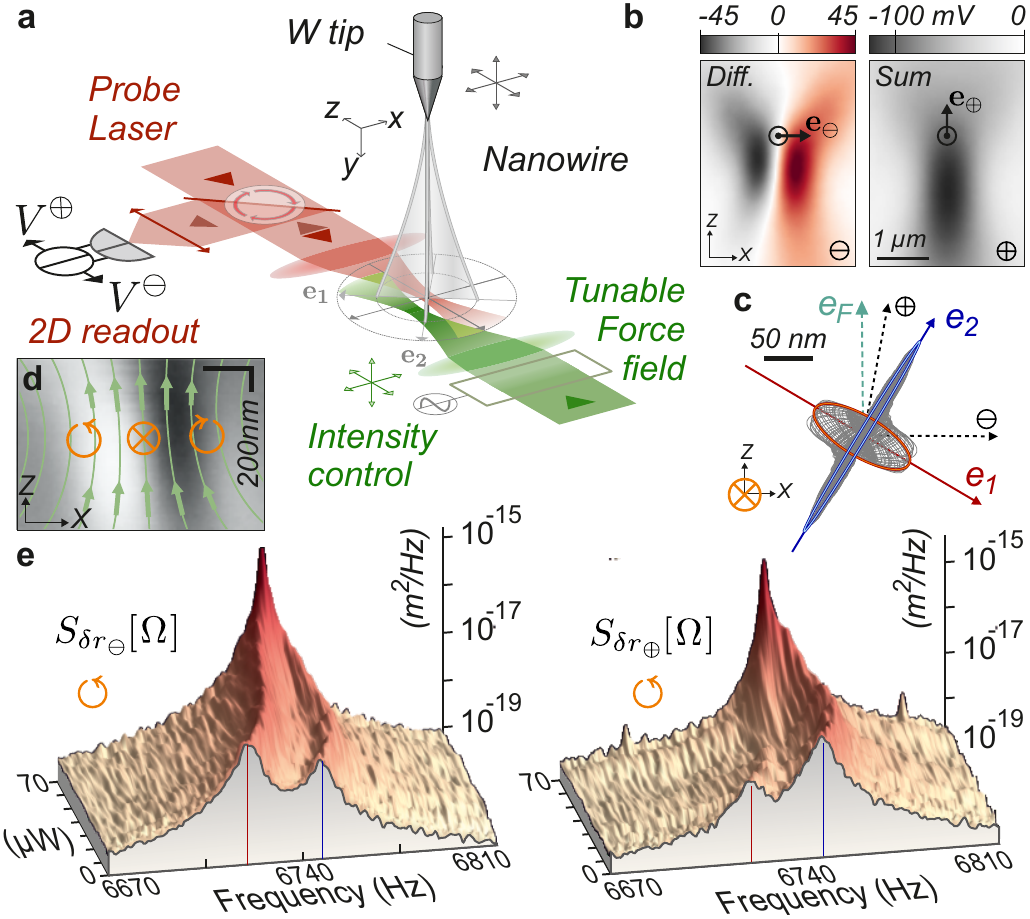}
\caption{
\textbf{ The experiment.} {\bf a} The transverse vibrations of a singly-clamped nanowire are optically read out by focusing a red probe laser beam on its vibrating extremity while recording the reflected intensity fluctuations on a dual photodetector. A counter propagating focused green laser beam generates a tunable optical force field of non-conservative nature \cite{Gloppe2014}.  {\bf b}  xz maps of the sum/difference DC signals $V_{\oplus/\ominus} (\vv{r})$ obtained by piezo-scanning the nanowire in the probe laser waist. The measurement location $\odot$ provides quasi-orthogonal projective measurement vectors $\vv{e_{\ominus,\oplus}}$ allowing a full reconstruction of the nanowire trajectories in 2D, $\vv{\delta r (t)}$. {\bf c} Steady-state driven trajectories obtained under resonant optical actuation at low optical power ($2\,\rm \mu W$) for a set of driving frequencies sampling both eigenmodes responses (300 curves from 6670 to 6810 kHz).  $\vv{e_F}$ : optical actuation force vector, $\vv{e_{1,2}}$: uncoupled eigenmodes orientations. {\bf d} Sketch of the optical force field structure superimposed on the differential green reflection image. Measurement positions indicated by $\otimes,\circlearrowleft,\circlearrowright$ feature zero and opposite force field vorticities respectively.  {\bf e} Calibrated thermal noise spectra measured on each measurement channels: $S_{\delta r_{\ominus,\oplus}}[\Omega]$ at  position $\circlearrowleft$ for increasing green optical power.
}
\label{Fig1}
\end{center}
\end{figure}
Similar nanoresonators were employed in MRFM applications \cite{Nichol2008,Nichol2012}, in electrostatic force field sensing \cite{Siria2012,MercierdeLepinay2016,Rossi2016}, while all the developments presented here also apply to doubly clamped nanobeams oscillating in and out of plane \cite{Faust2012,Moser2013,Faust2013,Stapfner2013,Anetsberger2009}. Such nanowires were previously employed to map the optical force field generated at the waist of a strongly focused laser beam, using pump-probe measurements at low power in absence of distortion. This revealed its non-conservative nature through a direct measurement of its non-zero force rotational \cite{Gloppe2014}.
Recently a novel universal method was introduced \cite{MercierdeLepinay2016} to image any 2D force fields even when pump-probe measurements \cite{Gloppe2014, Rossi2016} cannot be employed. Simultaneously tracking the eigenfrequency shifts and eigenmodes rotations of a quasi-frequency degenerated nanowire gives access to  the entire structure of the 2D force field gradient experienced by the nanowire extremity. In particular this nano-compass like method gives access to shear components which are essential to understand the physics at play in strongly varying force fields. The measurement principle was verified on a conservative electrostatic force field \cite{MercierdeLepinay2016}, this work proves its validity in the case of non-conservative force fields.\\

{\it The experiment--}
The experiment is conducted on a $165\,\rm \mu m$-long and  $\simeq 120\,\rm nm$-diameter Silicon Carbide nanowire (NW) suspended at the extremity of a sharp tungsten tip. Using high numerical aperture microscope objectives and a XYZ piezostage, a 633\,nm probe laser is strongly focused on the NW vibrating extremity whose position fluctuations $\vv{\delta r}(t)$ are encoded on the reflected field. It is collected on a split photodetector followed by a low noise amplifier providing the sum and difference of photovoltages: $V_{\ominus,\oplus} (\vv{r_0}+ \vv{\delta r}(t))$. Their temporal fluctuations $\delta V_{\ominus,\oplus} (t)= \vv{\delta r}(t)\cdot {\boldsymbol{\nabla} V_{\ominus,\oplus}} $ convey a projective measurement of the 2D NW trajectory,  $\delta r_{\ominus,\oplus}\equiv \vv{\delta r}(t)\cdot \vv{e_{\ominus,\oplus}}$,  projected along a  measurement vector  $\vv{e_{\ominus,\oplus}}\equiv {\boldsymbol{\nabla} V_{\ominus,\oplus}}/\left| {\boldsymbol{\nabla} V_{\ominus,\oplus}}\right |$, see Fig.\,1b. Operating at the working point highlighted by $\odot$ where both measurement vectors are quasi-orthogonal permits to realize a fully 2D readout of the NW position fluctuations through a simultaneous acquisition of both signals. A position tracking can be activated to stabilize the NW position with respect to the probe beam.\\
The non-conservative force field \cite{Gloppe2014} is generated by an independent counter-propagating tightly focused  532\,nm laser beam. The spatial structure of the optical force field is adjusted by moving the focusing objective with a XYZ piezo-scanner while its magnitude is varied with an acousto-optic modulator (AOM). It can also simultaneously be used to create an intensity modulation to drive the NW resonantly to probe its mechanical response (see SI). \\
{\it Intrinsic nanowire properties--}
They were determined in absence of green light, first by acquiring the noise spectral densities $S_{\delta r_{\ominus,\oplus}}[\Omega]$ measured on each projected displacement channels on two spectrum analyzers, see the foreground of Fig.\,1e. Both fundamental longitudinal eigenmodes are visible, oscillating around 6.7\,kHz with a quality factor of $\approx 3000$  identical on both modes within $5\%$ in vacuum ($10^{-3}\,\rm mbar$). They present a  quasi-degenerated character ($0.3\%$) with a frequency splitting of $(\Omega_2-\Omega_1)/2\pi \approx 20\,\rm Hz$, which renders the NW extremely sensitive to shear components of any external force field \cite{MercierdeLepinay2016}. The intrinsic eigenmodes' orientations ($\vv{e_{1,2}}$) are determined from the comparison of the peak spectral densities measured on each measurement channels. They form an angle of respectively -28 and +62 degrees with the $\vv{e_x}$ axis ($\approx 2^\circ$ uncertainty) and are thus found to be perfectly perpendicular in absence of external force field. The impact of the probe beam on the NW dynamics was suppressed by reducing optical force field gradients (operation with an enlarged optical waist at low power $15\,\rm \mu W$ on the optical axis, where any possible residual force is irrotational).
Their effective mass amounts to $M_{\rm eff}=1.5\times 10^{-15}\,\rm kg$, with a spring constant of $\approx 3\,\rm \mu N/m$, conveying a force sensitivity of $\approx 10\,\rm aN/\sqrt{Hz}$ at room temperature.\\
Response measurements were subsequently realized by injecting a small amount of green light (2$\,\rm \mu W$), which was intensity modulated with the AOM to exert a time-modulated force. After a transitory period the monochromatic drive generates a steady-state trajectory in space, whose projections along $\vv{e_{\ominus,\oplus}}$ are measured by the dual 2D readout  on a network analyzer featuring  two synchronized measurement channels. Sweeping  the excitation tone across both eigenfrequencies permits after a geometrical reconstruction to build the set of steady-state driven trajectories in 2D, see SI and Fig.\,1c. On the green optical axis, the optical actuation force vector is quasi-aligned with $\vv{e_z}$ and can thus drive both eigenmodes. This representation offers a very straightforward determination of the eigenmodes orientations in space, as directions where maximal oscillation amplitudes are observed, which are in excellent agreement with the above measurements derived from 2D thermal noise analysis. The linearity of the response was verified up to large driven amplitudes, approaching 300\,nm. Beyond, the NW may exit the linear measurement area while mechanical bistability appears at even larger drive.\\
\begin{figure*}[t!]
\begin{center}
\includegraphics[width=0.98\linewidth]{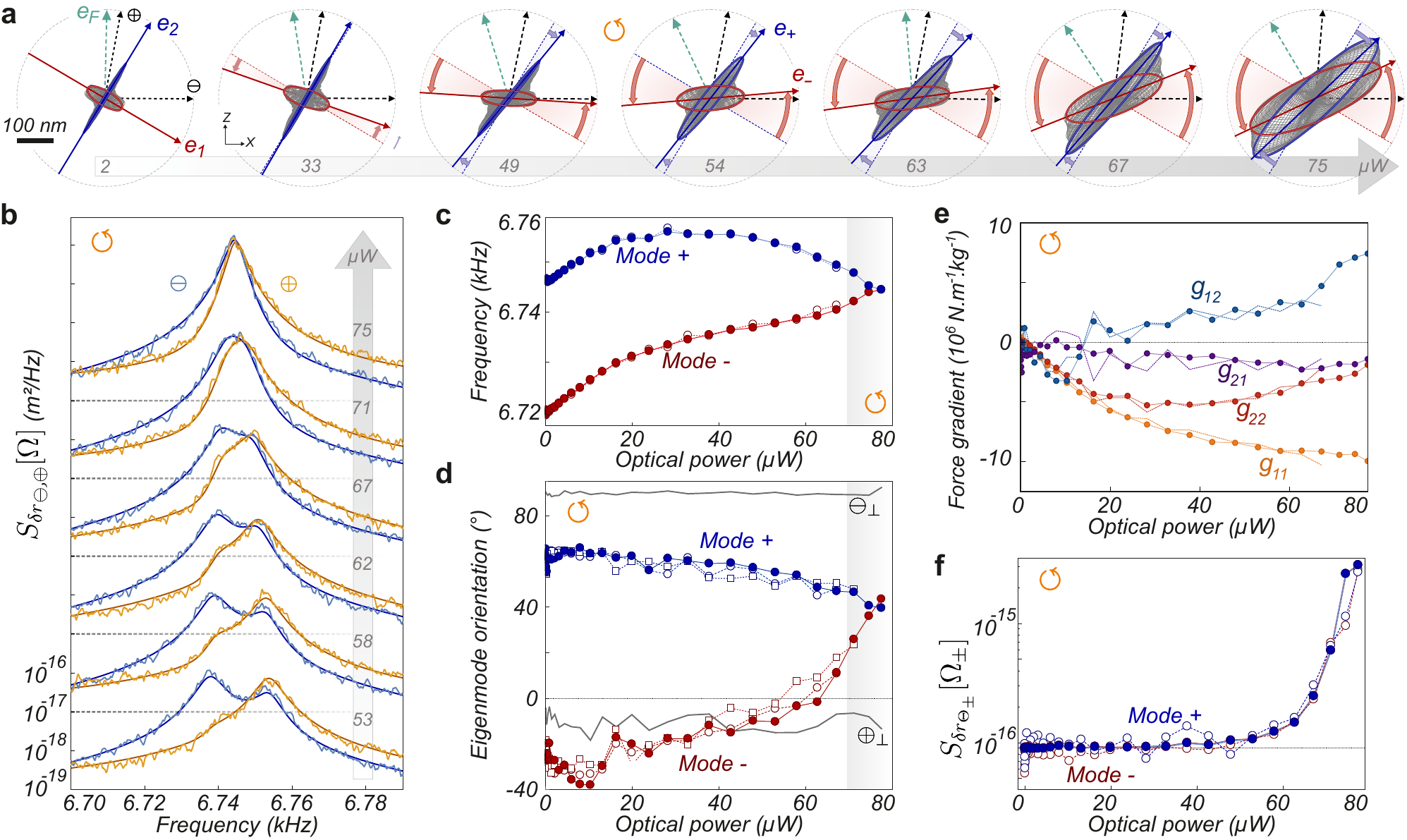}
\caption{
\textbf{ Eigenmode warping.} {\bf a} Driven trajectories in the xz plane for increasing green power but constant driving amplitude. The resonantly driven trajectories at  $\Omega_{+/-}$ are highlighted in blue/red. Eigenmodes orientations $\vv{e_\pm}$ loose their initial orthogonality and converge towards each other. {\bf b} Thermal noise spectra measured for increasing green power (with a 2 decade vertical offset between each set). Solid lines are fits using the model. {\bf c,d} Evolution of eigenfrequencies $\Omega_\pm/2\pi$ and eigenmode orientations $\theta_\pm$ with the green optical power. Angles derived from resonant analysis \cite{MercierdeLepinay2016} of thermal noise spectra/ driven trajectories  are marked as open circles/squares. Filled symbols are deduced from thermal noise spectral analysis (see text). {\bf e} Force field gradients deduced from thermal noise spectral analysis (full circles) and from the determined eigenfrequency shifts and eigenmode rotations shown above. {\bf f} Noise spectral densities $S_{\delta r_{\theta_\pm}} [\Omega_\pm]$ deduced from thermal noise analysis (open circles) and from the model, eq.\,(\ref{eq.sxres}), using the fitted force field gradients (filled circles).
}
\label{Fig2}
\end{center}
\end{figure*}
{\it Observation of eigenmode warping--}
We then immersed the NW in a non-conservative force field by increasing the green optical power and positioning the NW on the side of the green optical waist \cite{Gloppe2014} where the force field rotational is maximal  (indicated by $\circlearrowleft$ in Fig.\,1d).
Thermal noise spectra (Fig.\,1e,\,2b) and driven trajectories (Fig.\,2a) were recorded for increasing green optical powers, up to  $80\,\rm \mu W$  to avoid the topological instability \cite{Gloppe2014}. Strikingly visible on the 2D representations of the driven trajectories (Fig.\,2a), a strong warping of eigenmodes orientations is observed, which progressively rotate towards a common orientation pointing here around $+40^\circ$ from $\vv{e_x}$. The corresponding thermal noise spectra are shown in Fig.\,2b. They show  1) an eigenfrequency merging 2) a global noise increase and 3) anomalous spectral lineshapes. Eigenmodes orientations $\theta_\pm$ and eigenfrequencies $\Omega_\pm/2\pi$ deduced from both response and noise analysis are reported on Fig.\,2c and 2d.  Fig.\,2e shows the noise spectral densities $S_{\delta r_{\theta_\pm}}[\Omega_\pm]$ corresponding to the  displacement noise at the dressed eigenfrequencies computed for a projective measurement angle aligned with the eigenmode \cite{MercierdeLepinay2016}. A noise excess larger than 30 is observed when approaching the bifurcation. A factor of 2 only is expected for the independent summation of two uncorrelated noise spectral densities of eigenmodes with equivalent masses, frequencies, orientations, damping rates and temperatures. Radiation pressure noise is largely negligible here. A significant increase of the driven response is also visible in Fig.\,2a, despite the constant drive strength employed and the quasi-perpendicular orientation of the driving vector with respect to the dressed eigenmodes orientations close to the bifurcation. All these puzzling observations call for an in-depth investigation of the physics at play in the system.\\
\newsavebox{\smlmat}
\savebox{\smlmat}{$\left(\begin{smallmatrix}\Omega_1^2&0\\
0& \Omega_2^2\\
\end{smallmatrix}\right)$}
{\textit Model--} The dynamics of the NW deflection $\vv{\delta r} (t)$ around its rest position $\vv{r_0}$, restricted to the 2 fundamental eigenmodes follows:
$
\vv{ \delta \ddot r}=
- \boldsymbol{\Omega^2}\cdot \vv{\delta r}
-\Gamma \vv{ \delta \dot r}
+\frac{ \vv{ F}({\bf r_0}+\vv{ \delta r})+\vv{\vv{\delta F} + \delta F_{\rm th}}}{M_{\rm eff}}
$. $\boldsymbol{\Omega^2}\equiv$~\usebox{\smlmat} is the intrinsic restoring force matrix expressed in the unperturbed $\vv{e_{1,2}}$ basis, $\Gamma$  the mechanical damping rates, $M_{\rm eff}$  the effective masses, $\vv{\delta F}$ an external probe force and  $\vv{\delta F_{\rm th}}$ represents the  Langevin force vector which independently drives the NW \cite{Pinard1999} with a white force noise of spectral density $S_F^{\rm th}= 2 M_{\rm eff}\Gamma k_B T$ along each uncoupled axis.  Response measurements confirm the absence of delay on mechanical time scales in the establishment of the optical force experienced by the NW consecutive to an intensity change, as expected for radiation pressure forces \cite{Gloppe2014}. As a consequence, the external force experienced by the NW extremity  $\vv{F}(\vv{r_0}+\vv{\delta r})$ only depends on its position within the force field. For small position fluctuations with respect to the characteristic length scale of the force field structure, it can be linearized as $ \vv{F}(\vv{r_0})+\left.\left(\vv{\delta r}\cdot\boldsymbol{\nabla}\right)\vv{F}\right|_{\vv{r_0}}$. The static force causes a static NW deflection which can amount to $\approx 100\,\rm nm$  for the largest power employed and redefines the measurement position. The second linear term  represents an additional restoring force -proportional to the NW deflection- which modifies its dynamics. In Fourier space, we have $\vv{\delta r}[\Omega]= \boldsymbol{\chi}[\Omega]\cdot \vv{\delta F_{\rm th}}[\Omega]$
where $\boldsymbol{\chi}[\Omega]$  is the modified mechanical susceptibility matrix:
\begin{equation}
\boldsymbol{\chi^{-1}}[\Omega]\equiv M_{\rm eff} \left(
\begin{array}{cc}
{\tilde{\chi}}_1^{-1}[\Omega]-  g_{11} & - g_{21}\\
-g_{12}                                & {\tilde{\chi}}_2^{-1}[\Omega]-  g_{22}\\
\end{array}\right).
\label{eq.chiinv}
\end{equation}
$M_{\rm eff}^{-1}\tilde{\chi}_{1,2}[\Omega]\equiv 1/M_{\rm eff}(\Omega_{1,2}^2-\Omega^2-i\Omega \Gamma)$  are the original 1D mechanical susceptibilities. The NW dynamics now depends on the 4 components of the 2D external force field gradients:
$ g_{ij} (\vv{r_0})\equiv\frac{1}{M_{\rm eff}} \left.\partial_i F_j\right|_{\vv{r_0}} $
whose shear components ($i\neq j$) control the cross-coupling between eigenmodes.
Diagonalizing the susceptibility matrix gives the new eigenmodes, labeled with $\pm$ indices with eigenfrequencies:
$
\Omega_{\pm}^2\equiv\frac{\Omega_{1\parallel}^2+  \Omega_{2\parallel}^2}{2}\pm
\frac{1}{2}\sqrt{
\left(\Omega_{2\parallel}^2-\Omega_{1\parallel}^2\right)^2+ 4g_{12} g_{21}
}
$
and unitary eigenvectors
\newsavebox{\toto}
\savebox{\toto}{$\left(\begin{smallmatrix}\Delta \Omega_\perp^2\\ g_{12}\\
\end{smallmatrix}\right)$}
\newsavebox{\tata}
\savebox{\tata}{$\left(\begin{smallmatrix}-g_{21}\\ \Delta \Omega_\perp^2\\
\end{smallmatrix}\right)$}
${\vv{ e_-}}\equiv \frac{1}{\sqrt{g_{12}^2+\Delta \Omega_\perp^2}}$~\usebox{\toto} and
${\vv{ e_+}}\equiv \frac{1}{\sqrt{g_{21}^2+\Delta \Omega_\perp^2}}$~\usebox{\tata}. We used $\Omega_{i\parallel}^2\equiv\Omega_{i}^2-g_{ii}$ and $\Delta \Omega_\perp^2\equiv \Omega_{2\parallel}^2-\Omega_-^2=\Omega_+^2-\Omega_{1\parallel}^2$.
At first order shear components are responsible for eigenmode rotation and the angle between eigenvectors follows $\vv{e_{-}}\cdot\vv{e_+}\propto \vv{rot}(\vv{F})\cdot \vv{e_y}\propto g_{12}-g_{21}$. In a conservative force field, $g_{12}=g_{21}$, both eigenmodes are equally rotated and preserve their original orthogonality.  Instead, eigenmode orthogonality is broken in non-conservative force fields as experimentally observed.\\
From these expressions, see SI, one can compute projected thermal noise spectra $S_{\delta r_{\beta}}[\Omega]$ and steady-state trajectories  $\vv{\delta r} (t)= {\rm Re} \left(\boldsymbol{\chi}[\Omega]\, \delta F\, \vv{e_F} \, e^{-i\Omega t}\right)$ driven by a time-modulated force vector $\delta F \vv{e_F}$.\\
{\it Anomalous thermal noise spectra--}
Well visible in Fig.\,2b is the distortion of the measured thermal noise spectra into non-standard lineshapes presenting asymmetric peaks at large power. Such deviations, absent in conservative force fields \cite{MercierdeLepinay2016}, illustrate a large deviation from the normal mode expansion: thermal noise spectra cannot be correctly described by two eigenmodes featuring Lorentzian mechanical susceptibilities driven by independent Langevin forces. Such a deviation may originate from the conservative coupling of two resonators with heterogeneous damping \cite{Yamamoto2001,Schwarz2016}, but it is not the case here since no damping modification is measured all across the bifurcation.  Instead, each couple of thermal noise spectra can be very well fitted simultaneously with the model (solid lines in Fig.\,2b). The cold mechanical properties ($\Omega_{1,2}, \Gamma,\theta_{1,2}$ and $M_{\rm eff}$) were fixed to the previously determined values, so that the only fitting parameters are the 4 components of the force field gradients $g_{ij}$ which are reported in Fig.\,2e. As expected they vary with the green optical power, the observed deviation from ideal linearity is due to the static deflection which displaces the NW in the force gradient landscape. Since $g_{12}\neq g_{21}$, this directly demonstrates the non-conservative nature of the optical force field created by the green laser at the measurement position.\\
The above determination of the force field gradients results from fits of the entire NW thermal noise spectra in 2D. They can also be determined using the measured perturbation of the eigenmodes (rotation and frequency shifts) following the protocol exposed in \cite{MercierdeLepinay2016} and SI. They are reported in Fig.\,2e and a good agreement is observed between both methods. Interestingly, spectral analysis still operates close to the bifurcation while it becomes difficult to determine eigenmodes properties when they are not spectrally resolved anymore.\\
{\it Excess of thermal noise--}
Fig.\,2f shows the evolution with increasing pump power of $S_{\delta r_{\theta_\pm}}[\Omega_\pm]$, the thermal noise measured at resonance in the optimum readout configuration for a measurement vector aligned with the eigenmode. The model predicts  (see SI) a noise spectral density of:\\
\begin{equation}
S_{\delta r_{\theta_\pm}}[\Omega_\pm]= \frac{2 k_B T}{M_{\rm eff}\Gamma \Omega_\pm^2}\left(1+\frac{\left(g_{21}-g_{12}\right)^2}{\left(\Omega_+^2-\Omega_-^2\right)^2+\Gamma^2\Omega_\pm^2}\right)
\label{eq.sxres},\\
\end{equation}
and  thus a noise excess governed by $(\rm rot F )^2$. The 30-fold increase of the resonant noise spectral densites is well fitted by the above expression when using the experimentally derived force field gradients, see Fig.\,2f, which further confirms the validity of the model. Since a single 1D measurement cannot discriminate a noise increase from an eigenmode rotation, we underline that noise thermometry of 2D coupled systems can be subject to large misinterpretations if a proper 2D readout is not implemented.\\
\begin{figure}[t!]
\begin{center}
\includegraphics[width=0.99\linewidth]{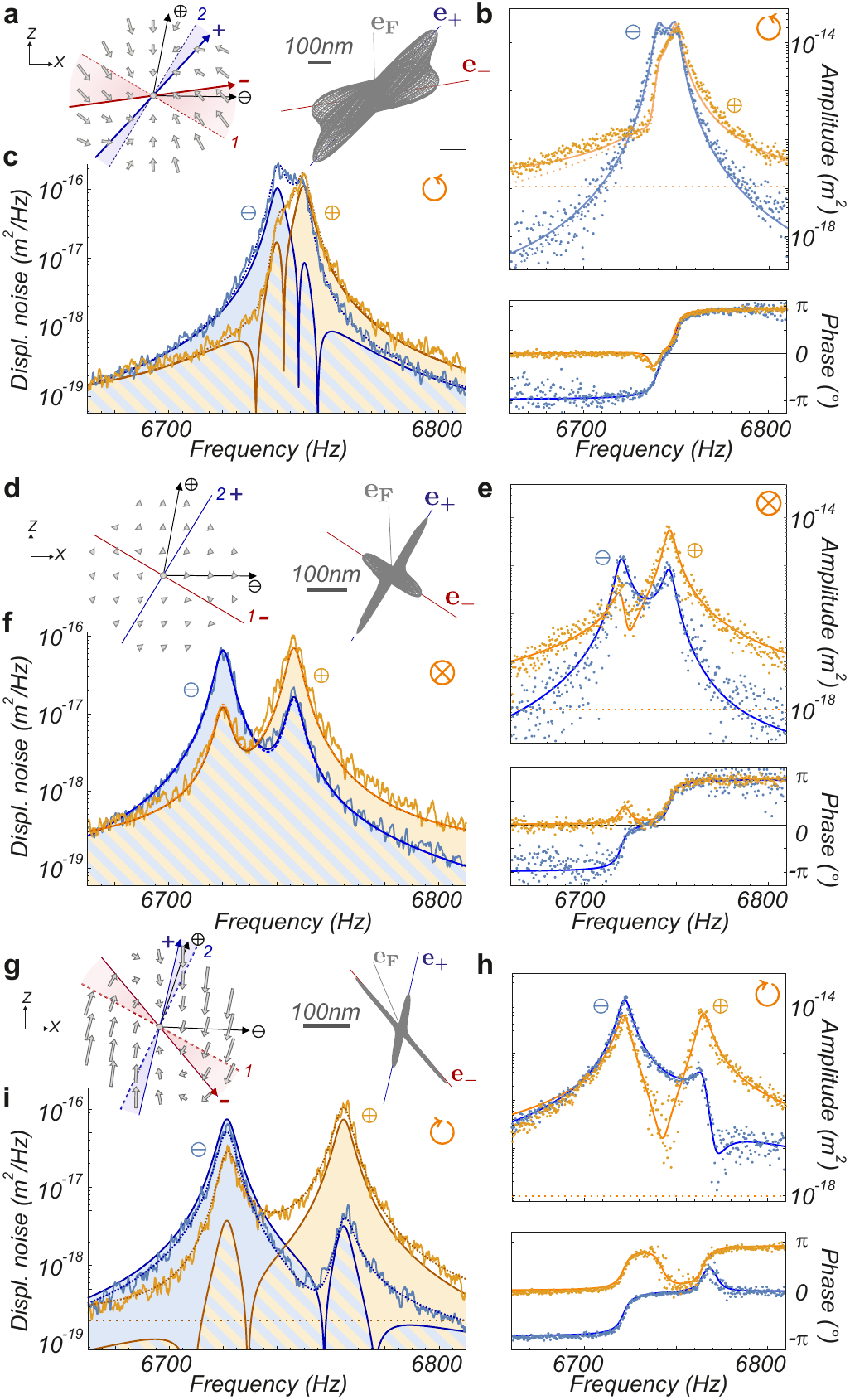}
\caption{
\textbf{ Deviation from the fluctuation-dissipation theorem.} {\bf a} Spatial structure of the optical force field at $\circlearrowleft$ built from the force gradients deduced from thermal noise analysis. The 2D representation of the driven trajectories are built from the response measurement obtained on each readout channels ($\ominus,\oplus$) shown in {\bf b}. They are fitted using the same $g_{ij}$ parameters  with an actuation vector indicated in the inset (270\,aN magnitude). {\bf c} Corresponding thermal noise spectra measured on each measurement channels, fitted with the above mentioned parameters (dashed lines). Solid lines are thermal noise spectra expected from the FDT, using expression (\ref{eq.FDT}) and the uniaxial mechanical susceptibilities $\chi_{\ominus\ominus}$, $\chi_{\oplus\oplus}$ respectively (see text) derived from response measurements. A pronounced  deviation  ($> 10\,\rm dB$) from experimental traces is observed. Panels {\bf d,e,f} are taken in position $\otimes$ at low optical power ($2\,\rm {\mu W}$) where the shear force field gradients can be neglected. There the measured spectra are in excellent agreement with the prediction of the FDT theorem. Panels {\bf g,h,i} are taken on the other side of the optical axis ($\circlearrowright$) with an opposite rotational where no frequency merging occurs, see SI. A similar substantial deviation is observed.
}
\label{Fig3}
\end{center}
\end{figure}
{\it  Deviation from the fluctuation-dissipation theorem--}
The strong asymmetry observed in the thermal noise spectra,  the obvious deviation from the normal mode expansions and the measured excess of noise naturally calls for a verification of the FDT.
For an oscillator at thermal equilibrium in the linear response regime, it connects the thermal noise spectrum $S_{\delta r_\beta}[\Omega]$ measured in an arbitrary direction $\vv{e_\beta}$ to the imaginary part of the NW mechanical susceptibility \cite{Callen1951,Kubo1966,Pinard1999} according to:
\begin{equation}
S_{\delta r_\beta}^{\rm FDT}[\Omega]= \frac{2k_B T}{|\Omega|} \left|{\rm Im} \, \chi_{\beta\beta} [\Omega]\right|,
\label{eq.FDT}
\end{equation}
where we have used the tensorial form of the 2D mechanical susceptibility: $\chi_{\mu\nu}[\Omega]\equiv \vv{e_\mu}\cdot{\boldsymbol \chi}[\Omega]\cdot\vv{e_\nu}$ obtained while driving along $\vv{e_\nu}$ and measuring along $\vv{e_\mu}$. The mechanical susceptibility involved thus corresponds to the one deduced from a response measurement realized with a test force having the same spatial profile as the readout mode \cite{Pinard1999}, oriented along the measurement vector $\vv{e_\beta}$.\\
Measurements were first performed on the left side of the optical axis (position $\circlearrowleft$ in Fig.\,1d). The orientation and magnitude of the driving force vector $\delta F \vv{e_F}$ are determined as fit parameters of the response measurements with the mechanical susceptibility of equation (\ref{eq.chiinv}) using the force field gradients ($g_{ij}$) deduced from the fit of the 2D Brownian motion. The results for both measurement channels  (both in amplitude and phase) are shown in Fig.\,3b and are in very good agreement with the data on a very large dynamical range ($> 40\,\rm  dB$) for a driving force of 270\,aN oriented at 94° degrees with respect to $\vv{e_x}$. In this situation, the driving vector is quasi aligned with the $\vv{e_\oplus}$ readout vector, see Fig.\,3a. The mechanical susceptibility $\chi_{\beta\beta}$ can then be evaluated using the same expressions with a force magnitude of 1\,N oriented along the measurement vector and the same set of $g_{ij}$, this for each measurement channels ($\ominus, \oplus$). Note that a weak coherent parasitic background was added to fit the data to account for parasitic electrical modulation but not taken into account in the noise fit functions. The thermal noise spectra expected from the FDT were then reported in Fig.\,3c (full lines) and compared to the spectra measured on each measurement channels. While both approaches are in good agreement far from resonance, a strong deviation is observed in the vicinity of eigenfrequencies, larger than 10 dB on each channel.
This discrepancy \cite{Harada2005} demonstrates the deviation from the FDT in our system.\\
A similar measurement set was realized in absence of non-conservative force field using lower static green optical power but identical intensity modulation to maintain the same driving force, see Fig.\,3d- 3f. Here also, the only free parameters were the orientation and strength of the driving force vector used to probe the mechanical responses. Here an excellent agreement is obtained between the measured and predicted thermal noise spectra, which validates the procedure.\\
Finally, measurements were performed on the other side of the beam at position $\circlearrowright$ in Fig.\,1d in an opposite rotational where eigenmodes also loose their orthogonality but are repelled from each other without frequency merging, see SI. A large deviation from the FDT is similarly observed at large pump power ($80\,\rm \mu W$), see Fig.\,3i. This further underlines that the deviation is not due to the frequency merging but instead to the eigenmode orthogonality breaking induced by the non-conservative force field.\\
\begin{figure}[t!]
\begin{center}
\includegraphics[width=0.98\linewidth]{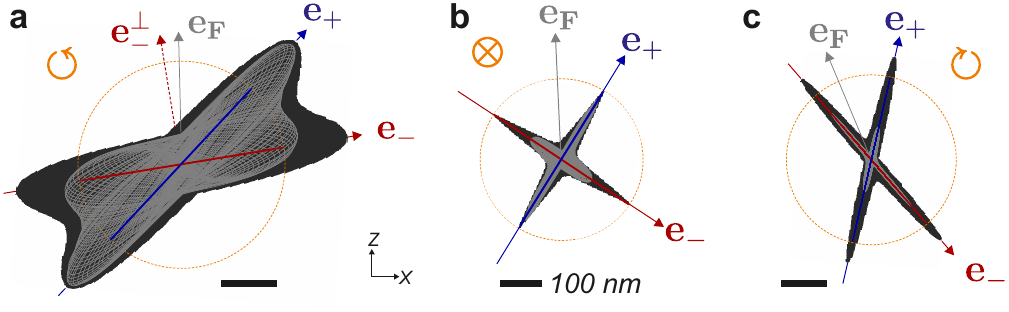}
\caption{
\textbf{  Tensorial structure of the 2D mechanical response.} Driven trajectories at the 3 different locations of Fig.\,3 for varying orientations of the drive vector (fixed magnitude of 270\,aN).  Red/blue curves are expected when driving along $\vv{e_{-/+}}$,  reaching a maximum amplitude represented as a dashed orange circle. Gray curves are measured for the drive orientation of Fig.\,3.  Black curves represent the entire set of trajectories obtained for all drive orientations, see SI.
}
\label{Fig4}
\end{center}
\end{figure}
{\it Non-axial contributions.}
Inspecting the geometrical properties of the tensorial susceptibility permits a phenomenological understanding of the deviation. We focus on the situation where the measurement vector is aligned with one eigenmode: $\vv{e_\beta}=\vv{e_+}$. A drive vector aligned with the eigenmode $\vv{e_F}=\vv{e_+}$  generates a driven displacement confined along the eigenmode and the tensorial susceptibility thus takes a uniaxial form:  $\chi_{++}= 1/M_{\rm eff}(\Omega_+^2-\Omega^2-i\Omega\Gamma)$. This is the one involved in the FDT theorem, valid in conservative force fields \cite{MercierdeLepinay2016} see SI and on the optical axis (Fig.\,3f). Its resonant value weakly differs from the uncoupled case ($<0.5\,\%$ relative frequency shifts) so the observed excess of noise cannot be explained by this uniaxial contribution and non-axial terms should now be inspected.
As shown in Fig.\, 4a (gray curves), a drive vector almost perpendicular to $\vv{e_-}$ can generate a displacement along $\vv{e_+}$, larger than under uniaxial driving (blue curves). Indeed, the maximum resonantly driven displacement measured along one eigenmode orientation is obtained for a drive exactly perpendicular to the other eigenmode: ${\theta^{\pm}_F}^{\rm opt}= \theta_\mp+\pi/2$, see SI. In absence of perpendicularity breaking, one recovers the intuitive result that the drive is most efficient when aligned with the eigenmode. As a consequence, in a non-conservative force field, the transverse components of the tensorial susceptibility  ($\chi_i^\perp\equiv \vv{e_i}\cdot \boldsymbol{\chi}\cdot \vv{e_i}^\perp$) allow each mode to be also driven by the second, lateral uncorrelated Langevin force.\\
This mechanism quantitatively accounts for the observed excess of noise. At any frequency, the bidimensional deviation verifies:
\begin{equation}
\begin{array}{ccl}
\sum\limits_{\mu=\beta,\beta^\perp}\left( S_{\delta r_\mu}- \frac{2 k_B T}{|\Omega|}\left| {\rm Im} \chi_{\mu\mu}\right|\right) &=& S_F \left|\chi_{12}-\chi_{21}\right|^2\\
 &=&  S_F \left|\chi_{-}^\perp \chi_{+}^\perp\right|\\
\end{array}
\end{equation}
which represents a patch for the FDT in our system, where the summation can be performed along any couple of perpendicular measurement orientations.  Its weighted spectral integral yields the power injected by the non-conservative force field, fully dissipated by damping forces (see SI), which is precisely a measure of the  FDT violation in out-of equilibrium systems according to the Harada-Sasa theorem \cite{Harada2005, Harada2006}. It also corresponds to the entropy production rate in the system \cite{Zamponi2005}, see SI, which further underlines from a thermodynamical point of view, the key role of the non-axial susceptibility. The  quadratic dependence of the deviation in $\rm \vv{rot}\vv{ F}$ translates the fact that the NW is exploring a rotational force field along Brownian trajectories whose statistics is unbalanced by the non-conservative force in an irreversible way.\\

\textit{Conclusions---}
We reported on eigenmodes warping, distortion of thermal noise spectra and  deviation from the FDT when immersing a multimode 2D nanoresonator in a non-conservative but non-dissipative force field. Our model quantitatively accounts for all observations and accurately describes the fluctuations and driven dynamics of the strongly and non-reciprocally coupled 2D system.  This work validates in the non-conservative case the principle of force field sensing \cite{MercierdeLepinay2016} based on recording of eigenmode orientation and frequency shifts. It underlines the subtlety arising in noise thermometry in strongly confined optical fields. Similar signatures should be observable in any non-reciprocally coupled dual physical systems and in particular in multimode cavity-optomechanics \cite{Schmidt2012,Massel2012,Seok2013,Shkarin2014,Buchmann2015,Xu2016} when the cavity can asymmetrize the mutual couplings. This system is also of great interest for testing advanced formulations of fluctuation theorems \cite{Seifert2012,Ciliberto2013, Gieseler2014,Prost2009} based on 2D trajectory analysis.\\

\textit{Acknowledgements---} We warmly thank the PNEC group at ILM, C. Elouard, A. Gloppe, F.\,Fogliano, L. Mougel,  J.P.\, Poizat, G.\, Bachelier, J.\,Jarreau, C.\,Hoarau, E.\,Eyraud and D.\,Lepoittevin. This project is supported by  ANR (FOCUS), ERC  (StG-2012-HQ-NOM) and Lanef (CryOptics).


\end{document}